\documentclass[twocolumn,amsmath,amssymb,floatfix,nofootinbib]{revtex4}

\usepackage{latexsym,amsmath,amssymb,theorem,dsfont}
\usepackage{amssymb} 
\usepackage{amsmath}
\usepackage{color}

\usepackage{amssymb}
\usepackage{amsmath}

\newcommand{\be}{\begin{equation}}
\newcommand{\ee}{\end{equation}}
\newcommand{\bea}{\begin{eqnarray}}
\newcommand{\eea}{\end{eqnarray}}
\newcommand{\beas}{\begin{eqnarray*}}
\newcommand{\eeas}{\end{eqnarray*}}
\newcommand{\nn}{\nonumber}

\def\lsim{\mathrel{\mathstrut\smash{\ooalign{\raise2.5pt\hbox{$<$}\cr\lower2.5pt\hbox{$\sim$}}}}}
\def\gsim{\mathrel{\mathstrut\smash{\ooalign{\raise2.5pt\hbox{$>$}\cr\lower2.5pt\hbox{$\sim$}}}}}

\definecolor{darkblue}{rgb}{0.15,0.35,0.55}
\definecolor{orangeish}{rgb}{0.65, 0.2, 0.2}
\usepackage[linktocpage=true]{hyperref}
\hypersetup{
colorlinks=true,
citecolor=darkblue,
linkcolor=orangeish,
urlcolor=darkblue,
pdfauthor={},
pdftitle={},
}

\begin{document}

\title{A Hidden Symmetry of the Galileon}

\author{Kurt Hinterbichler${}^{\rm a,}$\footnote{\tt khinterbichler@perimeterinstitute.ca} and Austin Joyce${}^{\rm b,}$\footnote{\tt ajoy@uchicago.edu} \\ \vspace{0.22in}}

\affiliation{${}^{\rm a}$Perimeter Institute for Theoretical Physics, 31 Caroline St. N., Waterloo, ON, N2L 2Y5, Canada \\ ${}^{\rm b}$Enrico Fermi Institute and Kavli Institute for Cosmological Physics, University of Chicago, Chicago, IL 60637, USA}

\begin{abstract}\noindent
We show that there is a special choice of parameters for which the galileon theory is invariant under an enhanced shift symmetry whose non-linear part is quadratic in the coordinates.  This symmetry fixes the theory to be equivalent to one with only even powers of the field, with no free coefficients, and accounts for the improved soft limit behavior observed in the quartic galileon S-matrix.
\end{abstract}

\maketitle 

%
%
Effective theories with derivative interactions have been of great interest recently.  Much of this work has focused on a particular family of scalar field theories, the galileons~\cite{Nicolis:2008in}. These theories have primarily been of interest in cosmology, where they arise in various infrared modifications of gravity (they appeared in the decoupling limit of the Dvali--Gabadadze--Porrati model~\cite{Dvali:2000hr,Luty:2003vm,Nicolis:2004qq}, and have since been seen to arise in massive gravity~\cite{deRham:2010ik,deRham:2010kj}), but they have potential applications in many corners of high energy physics, cosmology and even condensed matter.

Galileons appear in the non-relativistic limit of theories describing fluctuating hypersurfaces~\cite{deRham:2010eu,Goon:2011qf,Goon:2011uw,Burrage:2011bt}, which is of interest in high energy physics, but may also be of interest for biophysics or condensed matter applications. The galileons and their higher-shift analogues  may also be useful for describing Goldstone bosons near multi-critical points~\cite{Griffin:2013dfa,Hinterbichler:2014cwa,Griffin:2014bta}.

Galileons possess a number of interesting properties: they obey a non-renormalization theorem~\cite{Luty:2003vm,Hinterbichler:2010xn}, which indicates that they may be employed to address naturalness problems.  They can exhibit classically non-linear behavior without losing control of quantum corrections---this is the essence of the Vainshtein mechanism which can screen the presence of the galileon from Solar System tests~\cite{Vainshtein:1972sx,Babichev:2013usa}.  Additionally, the galileons can be interpreted as Wess--Zumino terms for a particular spontaneously broken space-time symmetry~\cite{Goon:2012dy}. For reviews of these properties along with cosmological applications and generalizations, see~\cite{Trodden:2011xh,deRham:2012az,Joyce:2014kja}.

The galileon has two essential defining properties: second order equations of motion (which ensures that the theory does not propagate an Ostrogradsky-type ghost) and invariance under the symmetry
\be
\delta\phi = c+b_\mu x^\mu,
\label{oldgalsymm}
\ee
where $c$ is a constant, $b_\mu$ is a constant vector and $x^\mu$ is the spacetime coordinate. There are a finite number of terms with these properties; $D+1$ of them in $D$ dimensions. 

In this letter, we show that, up to field redefinitions, there is a single choice of coefficients for the galileon terms for which the theory is additionally invariant under a higher-shift symmetry.  Up to field redefinitions, this symmetry fixes all the coefficients of the galileon, and the resulting theory is equivalent to one with only even powers of the field.  For example, in four dimensions, the theory containing only the quartic galileon term,
\be {\cal L}=-{1\over 2}(\partial\phi)^2+{1\over 12\Lambda^6}(\partial\phi)^2\Big[(\square\phi)^2-(\partial_\mu\partial_\nu\phi)^2\Big], \ee
where $\Lambda$ is the strong coupling scale,
is invariant under 
\be 
\delta \phi=s_{\mu\nu}x^\mu x^\nu+{1\over \Lambda^6} s^{\mu\nu}\partial_\mu\phi\,\partial_\nu\phi,\label{symm}
\ee
where $s_{\mu\nu}$ is a traceless symmetric constant tensor  $s_{\mu\nu}=s_{\nu\mu},~s^\mu_{\ \mu}=0.$

As with all global symmetries, this symmetry has consequences for correlation functions and scattering amplitudes.  We show that the soft-$\phi$ theorem associated to this extended shift symmetry implies that the soft limit of scattering amplitudes starts at ${\cal O}(q^3)$, higher than the  ${\cal O}(q^2)$ behavior of a generic galileon.  This explains a phenomenon seen recently in~\cite{Cheung:2014dqa}.  

\noindent
\textbf{Conventions:} We use mostly plus signature.  $D$ is the number of spacetime dimensions. $\simeq$ denotes equality up to total derivative.

\subsection{Galileon Lagrangians and Useful Quantities}
\vspace{-.2cm}

The galileon Lagrangians can be conveniently written in terms of certain total derivative combinations.
Define the matrix of second derivatives: $\Phi^{\mu}_{\ \nu}\equiv\partial^\mu\partial_\nu\phi$. At each order in $\phi$, there is a unique combination of $\Phi$'s that is a total derivative \cite{Creminelli:2005qk,Nicolis:2008in},
\begin{align}
\label{totalddef}
{\cal L}_n^{\rm TD}=\sum_p\left(-1\right)^{p}\eta^{\mu_1p(\nu_1)}&\eta^{\mu_2p(\nu_2)}\cdots\eta^{\mu_np(\nu_n)}\\\nonumber
&\times\left(\Phi_{\mu_1\nu_1}\Phi_{\mu_2\nu_2}\cdots\Phi_{\mu_n\nu_n}\right).
\end{align}
The sum is over all permutations of the $\nu$ indices, with $(-1)^p$ the sign of the permutation. 
The first few cases are
\bea {\cal L}_1^{\rm TD} &=&[\Phi]\nn\\
 {\cal L}_2^{\rm TD} &=&[\Phi]^2-[\Phi ^2] \nn\\
{\cal L}_3^{\rm TD}&=& [\Phi]^3-3 [\Phi][\Phi ^2]+2[\Phi ^3] \nn\\
{\cal L}_4^{\rm TD} &=&[\Phi]^4
-6[\Phi ^2][\Phi]^2+8[\Phi ^3][\Phi]+3[\Phi ^2]^2 -6[\Phi ^4]  \nn\\
&\vdots& 
\eea
where the brackets are traces of the enclosed matrix product.  We also define $ {\cal L}_0^{\rm TD}=1$.
Since one cannot anti-symmetrize more than $D$ indices in the definition \eqref{totalddef}, the term $ {\cal L}_n^{\rm TD}$ vanishes identically when $n>D$, so there are only a finite number of non-trivial such combinations.  

The galileon terms are given by
\be {\cal L}_n=-{1\over 2}(\partial\phi)^2{\cal L}_{n-2}^{\rm TD} \simeq {1\over n}\phi\, {\cal L}_{n-1}^{\rm TD},\label{lterms}\ee
with the last equality up to integration by parts \cite{Deffayet:2013lga}.  ${\cal L}_1=\phi$ is a tadpole and ${\cal L}_2 =  -{1\over 2}(\partial\phi)^2$ is the kinetic term.  The general galileon theory is a linear combination of the terms \eqref{lterms} with coefficients $c_1,\cdots,c_{D+1}$
\be {\cal L}=\sum_{n=1}^{D+1} c_n {\cal L}_n.\label{galillag}\ee
There is an energy scale, $\Lambda$, which suppresses the terms relative to each other, and at which the theory becomes strongly coupled.  We have chosen units such that $\Lambda=1$.   

An important ingredient will be the tensors $X^{(n)}_{\mu\nu}$ constructed out of $\Phi_{\mu\nu}$ as follows:\footnote{These are the same tensors which appear in the decoupling limit of massive gravity \cite{deRham:2010kj} (see the appendix of \cite{Hinterbichler:2011tt} for more on their properties).}
\be X^{(n)}_{\mu\nu}={1\over n+1}{\delta \over \delta \Phi_{\mu\nu}} {\cal L}_{n+1}^{\rm TD}.\label{Xtensors}\ee
The first few are
\bea X^{(0)}_{\mu\nu}&=&\eta_{\mu\nu}\nn \\
X^{(1)}_{\mu\nu}&=&\left[\Phi\right]\eta_{\mu\nu}-\Phi_{\mu\nu}\nn \\
X^{(2)}_{\mu\nu}&=&\left(\left[\Phi\right]^2-\left[\Phi^2\right]\right)\eta_{\mu\nu}-2\left[\Phi\right]\Phi_{\mu\nu}+2\Phi^2_{\mu\nu} \nn\\\nn
X^{(3)}_{\mu\nu}&=&\left(\left[\Phi\right]^3-3\left[\Phi\right]\left[\Phi^2\right]+2\left[\Phi^3\right]\right)\eta_{\mu\nu}\\
&&~-3\left(\left[\Phi\right]^2-\left[\Phi^2\right]\right)\Phi_{\mu\nu}+6\left[\Phi\right]\Phi^2_{\mu\nu}-6\Phi^3_{\mu\nu}\nn \\
&\vdots&
\eea
The $X^{(n)}_{\mu\nu}$ are symmetric, identically conserved $ \partial^\mu X^{(n)}_{\mu\nu}=0$, and satisfy the recursion relation \cite{deRham:2010kj}
\be X^{(n)}_{\mu\nu}=-n\,\Phi_\mu^{\ \lambda}X^{(n-1)}_{\lambda\nu}+{\cal L}_{n}^{\rm TD}\eta_{\mu\nu},\label{recursionsr}\ee
as well as the contraction property $ \Phi^{\mu\nu} X^{(n)}_{\mu\nu}={\cal L}_{n+1}^{\rm TD}.$ 

The most important property for what follows is that these tensors satisfy dimension dependent identities: $X^{(n)}_{\mu\nu}$ vanishes identically for $n\geq D$:
\be X^{(n)}_{\mu\nu}=0,\ \ \ \ n\geq D.\label{XvanishingD}\ee
This is because $ {\cal L}_n^{\rm TD}$ vanishes for $n>D$.

\subsection{The Symmetry}
\vspace{-.2cm}

The general transformation we will consider includes a part with no fields, a part with one power of the field, and a part with two powers of the field,
\be 
\delta\phi=\delta_0\phi+2\beta \delta_1\phi+  \left(\alpha+\beta^2\right) \delta_2\phi,
\label{shortsymmexp}
\ee
where $\alpha$, $\beta$ are constant parameters (the specific parameterization is chosen for later convenience), and 
\be \delta_0\phi=s_{\mu\nu}x^\mu x^\nu,\ \ \  \delta_1\phi=s_{\mu\nu}\partial^\mu\phi \,x^\nu,\ \ \ \delta_2\phi=s^{\mu\nu}\partial_\mu\phi\,\partial_\nu\phi,\label{symmpieces}\ee
with $s_{\mu\nu}$ a constant, symmetric, traceless tensor,
\be s_{\mu\nu}=s_{\nu\mu}~,\ \ \ s^\mu_{\ \mu}=0.\ee

The Euler--Lagrange derivatives of the variation of the general galileon terms under the three pieces \eqref{symmpieces} take a simple form in terms of the tensors~\eqref{Xtensors} (see the Appendix for the proof),
\begin{align}
\nonumber
\label{formual}
{\delta\over \delta\phi}\big(\delta_0 {\cal L}_n\big)&=2(n-1)s^{\mu\nu} X^{(n-2)}_{\mu\nu},\\
{\delta\over \delta\phi}\big(\delta_1 {\cal L}_n\big)&=-2s^{\mu\nu} X^{(n-1)}_{\mu\nu},\\ \nonumber
{\delta\over \delta\phi}\big(\delta_2 {\cal L}_n\big)&={2\over n}s^{\mu\nu} X^{(n)}_{\mu\nu}.
\end{align}

Looking at the form of \eqref{formual}, we see that if we choose relative coefficients properly, the terms of various order in the galileon Lagrangian \eqref{galillag} can be made to cancel against each other under the action of \eqref{shortsymmexp}, up to a total derivative.  To accomplish this, we demand
\be  c_n\left(\alpha+\beta^2\right)\delta_2 {\cal L}_n+2c_{n+1}\beta\delta_1 {\cal L}_{n+1}+c_{n+2}\delta_0{\cal L}_{n+2}\simeq 0,
\ee
which yields the recursion relation
\be \left(n + 1\right) c_{n + 2} - 
  2 \beta c_{n + 1} + \left(\alpha + \beta^2\right){1\over n} c_n = 0 .\label{recursionre}\ee
This relation determines all the coefficients of \eqref{galillag} in terms of $c_1,c_2$ and the parameters $\alpha,\beta$ of the transformation.

To establish invariance of the action, we must also show that the lowest order terms are invariant under the lowest order parts of the symmetry, and that the highest terms are invariant under the highest parts of the symmetry.  For the lower part, it is straightforward to see that
the kinetic term and tadpole terms are invariant up to a total derivative under the lowest order parts of the symmetry,
\be  \delta_0 {\cal L}_2 \simeq 0,~~~~~~\delta_0 {\cal L}_1 \simeq 0,~~~~~~\delta_1 {\cal L}_1\simeq 0.\ee
The highest terms are invariant under the higher order parts of the symmetry because of the dimension-dependent identity \eqref{XvanishingD}
\be \delta_2 {\cal L}_D \simeq 0,~~~~~\delta_2 {\cal L}_{D+1} \simeq 0,~~~~~\delta_1 {\cal L}_{D+1} \simeq 0.\ee

Without loss of generality, we may take $c_1=0$ by expanding around a background solution $\phi \propto x^2$  \cite{Nicolis:2008in} and, assuming the background is stable, we may canonically normalize the kinetic term to set $c_2=1$ (the form of the ansatz \eqref{shortsymmexp} is unchanged under these field redefinitions).  Taking these values as initial conditions, the recursion relation \eqref{recursionre} can be solved to give
\be c_n={\left(\beta+\sqrt{-\alpha}\right)^{n-1}-\left(\beta-\sqrt{-\alpha}\right)^{n-1} \over 2\sqrt{-\alpha}(n-1)!}.\label{recsolv}\ee

\subsection{Behavior Under Galileon Duality}
\vspace{-.2cm}

Galileon duality \cite{Fasiello:2013woa,deRham:2013hsa} gives a one-parameter redundancy of the galileon Lagrangians.  By performing a field redefinition
\be \phi'=e^{\theta\delta}\phi,\ \ \ \ \ \ \ \ \delta\phi=-{1\over 2}(\partial\phi)^2,\ee
we transform a galileon theory with one set of parameters into a galileon theory with a different set of parameters which are related by~\cite{Kampf:2014rka},
\be
\nonumber
\sum_{n=1}^{D+1} c_n{\cal L}_n(\phi')= \sum_{n=1}^{D+1} d_n(\theta){\cal L}_n(\phi),\  \ d_n(\theta)=\sum_{m=1}^n{\theta ^{n-m}\over (n-m)!}c_m.
\label{dualitycoefficients}
\ee
Under this duality, the symmetry~\eqref{symm} with $\phi\rightarrow\phi'$ becomes
\begin{align}
\delta \phi=&~ s_{\mu\nu}x^\mu x^\nu+2(\beta+\theta) s_{\mu\nu}x^\mu\partial^\nu\phi \nn\\
&~~~~~~+\left(\alpha+(\beta+\theta)^2\right) s^{\mu\nu}\partial_\mu\phi\,\partial_\nu\phi~.\label{dualsymm}
\end{align}

Given our canonically normalized theory with no tadpole, the coefficient $c_3$ simply shifts by $\theta$ under duality, so a convenient way to fix the duality ambiguity is to choose $\theta$ so that $c_3=0$.  From \eqref{recsolv}, we have $c_3=\beta$, so we choose $\theta=-\beta$, after which the symmetry \eqref{dualsymm} takes the form
\be \delta\phi=s_{\mu\nu}x^\mu x^\nu+\alpha \, s^{\mu\nu}\partial_\mu\phi\,\partial_\nu\phi~,\label{newsymmd} \ee
and the Lagrangian coefficients \eqref{recsolv} become
\be c_n={\left(-\alpha\right)^{{n\over 2}-1}\over  (n-1)!},\ \ \ ~~ n=2,4,6,\ldots\label{cnexpress}\ee 
with the odd $c_n$'s vanishing.

We see that once the duality ambiguity is removed, the theory with the symmetry~\eqref{newsymmd} contains only even powers of the field, with coefficients completely fixed in terms of one parameter $\alpha$, which, furthermore, can be re-absorbed into the energy scale $\Lambda$ by changing units.  Thus, the theory is completely fixed, with no free parameters other than the energy scale of strong coupling.

\subsection{Symmetry Algebra}
\vspace{-.2cm}

The generators of the galileon symmetry \eqref{oldgalsymm} are
\begin{align}
C\phi &= 1~,~~~~~~~{B^\mu}\phi = x^\mu~,~~~ 
\end{align}
Along with the standard  linear Poincar\'e generators ${P_\mu}\phi=-\partial_\mu\phi$, 
${J_{\mu\nu}}\phi=(x_\mu\partial_\nu-x_\nu\partial_\mu)\phi ,$ they close to form the galileon algebra \cite{Goon:2012dy}, whose non-zero commutators are 
\be
\nn
\left [P_\mu,B_\nu\right ] = \eta_{\mu \nu }C~,\ \ \ 
\left[J_{\mu\nu},B_{\lambda}\right] = \eta_{\mu\lambda }B_{\nu }-\eta_{\nu\lambda}B_{\mu}~,
\ee
 along with the standard commutators $\left [J_{\mu\nu },P_{\lambda }\right ] =\eta_{\mu\lambda}P_{\nu } - \eta_{\nu\lambda }P_{\mu }$,
$\left [J_{\mu\nu },J_{\lambda\sigma }\right ] = \eta_{\mu\lambda}J_{\nu\sigma }-\eta_{\nu\lambda }J_{\mu\sigma }+\eta_{\nu\sigma}J_{\mu\lambda }-\eta_{\mu\sigma}J_{\nu\lambda }$ of the Poincar\'e algebra.
 
There is a new symmetric traceless generator associated with the new symmetry \eqref{newsymmd}
\be
\nonumber
{ S_{\mu\nu}}\phi = x_\mu x_\nu - \frac{1}{D}x^2\eta_{\mu\nu}+\alpha \left[\partial_\mu\phi\,\partial_\nu\phi-\frac{1}{D}(\partial\phi)^2\eta_{\mu\nu}\right].
\ee
This generator closes with the galileon algebra to form an enlarged symmetry algebra whose new non-zero commutators are %
\begin{align}
\nn
\left[P_{\mu}, S_{\nu \lambda}\right] &= \eta_{\mu\nu}B_\lambda+\eta_{\mu\lambda}B_\nu-\frac{2}{D}B_\mu\eta_{\nu\lambda}~, \\
\left[B_{\mu}, S_{\nu \lambda}\right] & =-\alpha\Big( \eta_{\mu\nu}P_\lambda+\eta_{\mu\lambda}P_\nu-\frac{2}{D}P_\mu\eta_{\nu\lambda}\Big),  \\ \nonumber
\left [S_{\mu\nu },S_{\lambda\sigma }\right ] &=\alpha\left(\eta_{\mu\lambda}J_{\nu\sigma }+\eta_{\nu\lambda }J_{\mu\sigma }+\eta_{\nu\sigma}J_{\mu\lambda }+\eta_{\mu\sigma}J_{\nu\lambda }\right), \\ \nn
\left[J_{\mu\nu}, S_{\lambda\sigma}\right]& =\eta_{ \mu\lambda} S_{\nu\sigma}- \eta_{ \nu\lambda  } S_{ \mu\sigma }+\eta_{  \mu\sigma} S_{ \lambda \nu}-\eta_{  \nu\sigma} S_{ \lambda \mu}~.
\end{align}
When $\alpha\rightarrow 0$, this reduces to the algebra of traceless $N=2$ extended shift symmetries studied in \cite{Hinterbichler:2014cwa}.

\subsection{Soft Limit}
\vspace{-.2cm}

Recently, the authors of~\cite{Cheung:2014dqa} studied the behavior of soft limits of scattering amplitudes in various scalar field theories in $D=4$. In particular, they found that DBI has better behavior (with amplitudes scaling as ${\cal O}(q^2)$ in the soft limit) than a generic $P(X)$ theory, and that the general galileon has the best soft behavior among theories whose terms have $N$ fields and $2(N-1)$ derivatives. Additionally, they found a scalar field theory which has even better soft behavior, with its scattering amplitudes scaling as ${\cal O}(q^3)$ in the soft limit, which they conjectured to be the quartic galileon.  We are now in a position to see that this is a consequence of invariance under the extended shift symmetry \eqref{newsymmd}.

This symmetry leads to a ``soft-$\phi$" theorem of the following form (which is analogous to the soft pion theorems of chiral perturbation theory \cite{Adler:1964um,Weinberg:1966kf} or the soft-$\zeta$ theorems in cosmology \cite{Creminelli:2012ed,Assassi:2012zq,Hinterbichler:2013dpa}):
\be
\lim_{q\to 0} \partial_{q^{(\mu}}\partial_{q^{\nu)_T}}\left(\frac{\langle \phi_q \phi_{k_1}\cdots\phi_{k_N}\rangle'}{\langle\phi_q\phi_{-q}\rangle'}\right) = \hat{\cal D}\langle \phi_{k_1}\cdots\phi_{k_N}\rangle'~,
\label{wardid}
\ee
which says that the traceless part of the ${\cal O}(q^2)$ soft limit of the $(N+1)$-point correlation function is given by some differential operator $\hat{\cal D}$ acting on the $N$-point correlator (the prime denotes a correlation function without the momentum-conserving delta function). The precise form of $\hat{\cal D}$ is not important for our present purposes.  Now, our theory contains only galileons with even numbers of fields, and this ${\mathbb Z}_2$ symmetry causes all odd-point amplitudes to vanish. Therefore, if $(N+1)$ is even, the right hand side of the identity~\eqref{wardid} is zero, and~\eqref{wardid} tells us that the traceless part of the ${\cal O}(q^2)$ part of the soft limit vanishes. Since the galileon is massless, its $4$-momentum is null, so the trace part vanishes as well. Therefore, in the theory of the quartic galileon, the soft limit of amplitudes starts at ${\cal O}(q^3)$, in agreement with the findings of~\cite{Cheung:2014dqa} in explicit computations. Note that in higher dimensions, the theory \eqref{cnexpress} we have identified with this extended symmetry will also enjoy this improved soft limit behavior, as will all the theories related to it by galileon duality.  This is also the same special galileon theory for which an exact S-matrix was conjectured in \cite{Cachazo:2014xea}.  In $D=4$ it is the Legendre self-dual model described in \cite{Curtright:2012gx}.

\subsection{Conclusions}
\vspace{-.2cm}

We have identified a family of galileon theories which are invariant under an extended symmetry consisting of a shift quadratic in space-time coordinates and a shift quadratic in the field. The presence of this symmetry explains the soft behavior of scattering amplitudes in these theories. It is possible that this structure generalizes to higher shifts in both space-time coordinates and fields, which would lead to theories with even better soft behavior in higher dimensions. We note that a $0^{\rm th}$ order requirement that makes this plausible is that the kinetic term is invariant under an arbitrarily high-order traceless shift symmetry \cite{Hinterbichler:2014cwa}.


\subsection*{Appendix: Proof of \eqref{formual}\label{proofappendix}}
\vspace{-.2cm}
Write the galileon Lagrangians in the form ${\cal L}_n={(n-1)!\over n}\phi\, \Phi_{\mu_1}^{\ [\mu_1}\cdots \Phi_{\mu_{n-1}}^{\ \mu_{n-1}]}$, with anti-symmetrization of weight one, consistent with \eqref{lterms}, \eqref{totalddef}.  Using the anti-symmetrization of the derivatives, any variation can be written as
\be \delta {\cal L}_n\simeq { (n-1)!}\delta\phi \,\Phi_{\mu_1}^{\ [\mu_1}\cdots \Phi_{\mu_{n-1}}^{\ \mu_{n-1}]}.\ee
The variation under $\delta_0$ is 
\be \delta {\cal L}_n\simeq { (n-1)!}s_{\mu\nu}x^\mu x^\nu \Phi_{\mu_1}^{\ [\mu_1}\cdots \Phi_{\mu_{n-1}}^{\ \mu_{n-1}]},\ee
which upon integration by parts and using anti-symmetry becomes
\be \delta {\cal L}_n\simeq {2 (n-1)!}s^{\mu}_{\ \nu} \phi\, \delta_{\mu}^{\ [\nu}\Phi_{\mu_2}^{\ \mu_2}\cdots \Phi_{\mu_{n-1}}^{\ \mu_{n-1}]}.\ee
Taking the variation with respect to $\phi$, again using anti-symmetry, 
\bea  {\delta\over \delta\phi}\left(\delta_0 {\cal L}_n\right)& =&  {2(n-1) (n-1)!}s^{\mu}_{\ \nu} \delta_{\mu}^{\ [\nu}\Phi_{\mu_2}^{\ \mu_2}\cdots \Phi_{\mu_{n-1}}^{\ \mu_{n-1}]} \nn\\
&=&2(n-1)s^{\mu\nu} X^{(n-2)}_{\mu\nu},\eea
giving the first line of \eqref{formual}.

The variation under $\delta_2$ is 
\be
\delta_2 {\cal L}_n\simeq { (n-1)!}s^{\mu\nu}\partial_\mu\phi\,\partial_\nu\phi \,\Phi_{\mu_1}^{\ [\mu_1}\cdots \Phi_{\mu_{n-1}}^{\ \mu_{n-1}]},\ee
Taking the variation with respect to $\phi$, all the contributions with three or four derivatives on any $\phi$ cancel out, and what remains are the two-derivatives contributions,
\begin{align} 
& {\delta\over \delta\phi}\left(\delta_2 {\cal L}_n\right) =-2{ (n-1)!}s^{\mu\nu}\partial_\mu\partial_\nu\phi\, \Phi_{\mu_1}^{\ [\mu_1}\cdots \Phi_{\mu_{n-1}}^{\ \mu_{n-1}]} \nn \\
& +2(n-1){ (n-1)!}s^{\mu\nu}\partial_\mu\partial^{\lambda}\phi\,\partial_\nu\partial_{\sigma}\phi\, \delta_{\lambda}^{\ [\sigma}\Phi_{\mu_2}^{\ \mu_2}\cdots \Phi_{\mu_{n-1}}^{\ \mu_{n-1}]}  \nn\\
&=-2s^{\mu\nu}\left[\Phi_{\mu\nu}{\cal L}_{n-1}^{\rm TD} -(n-1)\Phi_{\mu}^{\ \lambda}\Phi_{\nu}^{\ \sigma}X^{(n-2)}_{\lambda\sigma}\right].\label{inte}
\end{align}
Now we use the recursion relation \eqref{recursionsr} twice in order to reduce the second term in the brackets:
\begin{align} 
& \Phi_{\mu}^{\ \lambda}\Phi_{\nu}^{\ \sigma}X^{(n-2)}_{\lambda\sigma}={1\over n-1} \Phi_{\mu}^{\ \lambda}\left[-X^{(n-1)}_{\nu\lambda}+{\cal L}_{n-1}^{\rm TD}\eta_{\nu\lambda}\right]\nn\\
&={1\over n-1} \left[{1\over n}\left(X^{(n)}_{\mu\nu}-{\cal L}_{n}^{\rm TD}\eta_{\mu\nu}\right)+\Phi_{\mu\nu}{\cal L}_{n-1}^{\rm TD}\right]. 
\end{align}
There is a cancellation between the final term here and the first term in the brackets of \eqref{inte}, and we may ignore the term proportional to $\eta_{\mu\nu}$ because of the tracelessness of $s_{\mu\nu}$, leaving the result in the last line of \eqref{formual},
${\delta\over \delta\phi}\left(\delta_2 {\cal L}_n\right)={2\over n}s^{\mu\nu} X^{(n)}_{\mu\nu}.$
The proof of the second line of \eqref{formual} follows similarly.

\noindent
{\bf Acknowledgments:}  We thank Olivier Gelling for pointing out some typos in v2.  Research at Perimeter Institute is supported by the Government of Canada through Industry Canada and by the Province of Ontario through the Ministry of Economic Development and Innovation.  This work was made possible in part through the support of a grant from the John Templeton Foundation. The opinions expressed in this publication are those of the author and do not necessarily reflect the views of the John Templeton Foundation (KH).
This work was supported in part by the Kavli Institute for Cosmological Physics at the University of Chicago through grant NSF PHY-1125897, an endowment from the Kavli Foundation and its founder Fred Kavli, and by the Robert R. McCormick Postdoctoral Fellowship (AJ).

\end{document}